\documentclass[twocolumn,prb,floatfix,unsortedaddress]{revtex4}
\usepackage{graphicx, subfigure, amsfonts,amssymb,amsmath, array, gensymb,fontenc,color}
\begin{document}

\title{Single-layer 1\emph{T'}-MoS\textsubscript{2} under electron irradiation from \emph{ab initio} molecular dynamics}

\author{Michele Pizzochero}
\email{michele.pizzochero@epfl.ch}
 
\author{Oleg V. Yazyev}
\email{oleg.yazyev@epfl.ch}
 
\affiliation{Institute of Physics,  Ecole Polytechnique F\'ed\'erale de Lausanne (EPFL), CH-1015 Lausanne, Switzerland}

\date{\today}

\begin{abstract}
Irradiation with high-energy particles has recently emerged as an effective tool for tailoring the properties of two-dimensional transition metal dichalcogenides. In order to carry out an atomically-precise manipulation of the lattice, a detailed understanding of the beam-induced events occurring at the atomic scale is necessary.  Here, we investigate the response of $1T'$-MoS$_2$ to the electron irradiation by \emph{ab initio} molecular dynamics means. Our simulations suggest that an electron beam with energy smaller than 75~keV does not result in any knock-on damage. The displacement threshold energies are different for the two nonequivalent sulfur atoms in $1T'$-MoS$_2$ and strongly depend on whether the top or bottom chalcogen layer is considered.  As a result, a careful tuning of the beam energy can promote the formation of ordered defects in the sample. We further discuss the effect of the electron irradiation in the neighborhood of a defective site, the mobility of the sulfur vacancies created and their tendency to aggregate. Overall, our work provides useful guidelines for the imaging and the defect engineering of $1T'$-MoS$_2$ using electron microscopy.
\end{abstract}


\maketitle

\section{Introduction} 

Two-dimensional transition metal dichalcogenides (TMDs) are a novel class of atomically thin crystals of the general formula $MX_2$, where the transition metal $M$ is sandwiched between two layers of chalcogen atoms $X$\cite{manz17}. In this large library of systems, group VI ($M$ = Mo or W) disulfides, diselenides and ditellurides have attracted considerable interest in the last years, mostly because of their rich polymorphism and peculiar electronic properties\cite{chh13}. Depending on the coordination around the metal atom, in fact, group VI single-layer TMDs can exist either in the stable $2H$ or in the metastable $1T'$ crystalline phase\footnote{Even though "2$H$" and "1$T'$" notation refers to three-dimensional bulk phases of layered transition metal dichalcogenides, throughout this work we adopt it also for the single-layer limit.}.

The combination of the reduced dimensionality with the presence of a direct band gap\cite{mak10, yaz15} have put the $2H$ phase of TMDs in the spotlight of emerging technologies such as flexible electronics and optoelectronics\cite{wan12}. Recent innovations include \emph{e.g.} the realization of field effect transistors\cite{rad11, mwchen17} and integrated circuits\cite{rad11bis} operating at room temperature, gas sensors for ammonia and humidity detection\cite{late13} or ultrasensitive photodetecting devices\cite{lop13}.  On the other hand, the semimetallic $1T'$ phase of TMDs has attracted attention in the community due to its topological electronic properties. It has been theoretically predicted that the tiny band-gap induced by the spin-orbit interactions together with the band inversion around the Fermi level lead to the quantum spin Hall (QSH) effect in this crystalline phase\cite{qian14, pulk17}. These predictions have been experimentally confirmed in the case of $1T'$-WTe$_2$\cite{tang17, fei17}. 

Among all transition metal dichalcogenides, MoS$_2$ is considered the most representative example. Recently, it has been found that a controlled introduction of defects in $2H$-MoS$_2$ can tailor many chemical and physical properties of the material\cite{lin16}. For instance, it has been shown that sulfur vacancies act as a catalytic centers for the hydrogen evolution reaction\cite{li16bis, ye16}, affect thermal\cite{wan16} and electron transport\cite{qiu13, pulkin16}, lead to magnetism when strained\cite{tao14} or passivated with metal elements\cite{siv16}, form ordered one-dimensional defects\cite{wan16bis, kom12} and offer a reactive site for the insertion of dopant species\cite{kom12}.  In addition, the introduction of such defects can be carried out in a controlled manner when the sample is exposed to an electron irradiation, thereby turning electron microscopy from a tool for material imaging to an effective strategy for an \emph{in situ} defects creation\cite{kom12, kom13, krash10, susi-2d17, zhao-mrs17}. Remarkably, it has been experimentally shown that electron beam exposure is able to promote a crystalline phase transition from the stable $2H$ to the metastable $1T'$, yielding to the realization of lateral semiconductor-semimetallic heterostructures between different polymorphs\cite{lin2014, katagiri16}. First-principles simulations have suggested that this transformation is triggered by the the disorder in the sample\cite{kretsch17}, highlighting once more the importance of lattice imperfections.

The large body of theoretical and experimental work mentioned above has focused on electron irradiation of the $2H$-phase MoS$_2$, while its effect on the $1T'$ phase was not addressed so far. We fill this gap in the following by exploring for the first time the response of $1T'$-MoS$_2$ to the electron beam based on \emph{ab initio} molecular dynamics.  We discuss the atomic structure of the defects that form upon electron irradiation at different beam energies, their mobility through the lattice and their possible tendency to aggregate. Furthermore, we conduct a comparison between the displacement threshold energy, the key quantity that governs the knock-on damage under the beam, and the formation energy, that accounts for the relative stability of defects under thermodynamic equilibrium conditions.  Our findings provide useful insights for experimental researchers by identifying the range of beam energies needed  to carry out $1T'$-MoS$_2$ imaging without inducing any radiation damage in the sample as well as guidelines for the controlled introduction of defects in the transmission electron microscope (TEM).

The rest of this paper is organized as follows. Section II presents the methodology adopted throughout this work together with the details of  first-principles calculations. In Section III we discuss the results of our simulations. Finally, Section IV briefly summarizes and concludes our work.

\begin{figure}
  \centering
  \includegraphics[width=1\columnwidth]{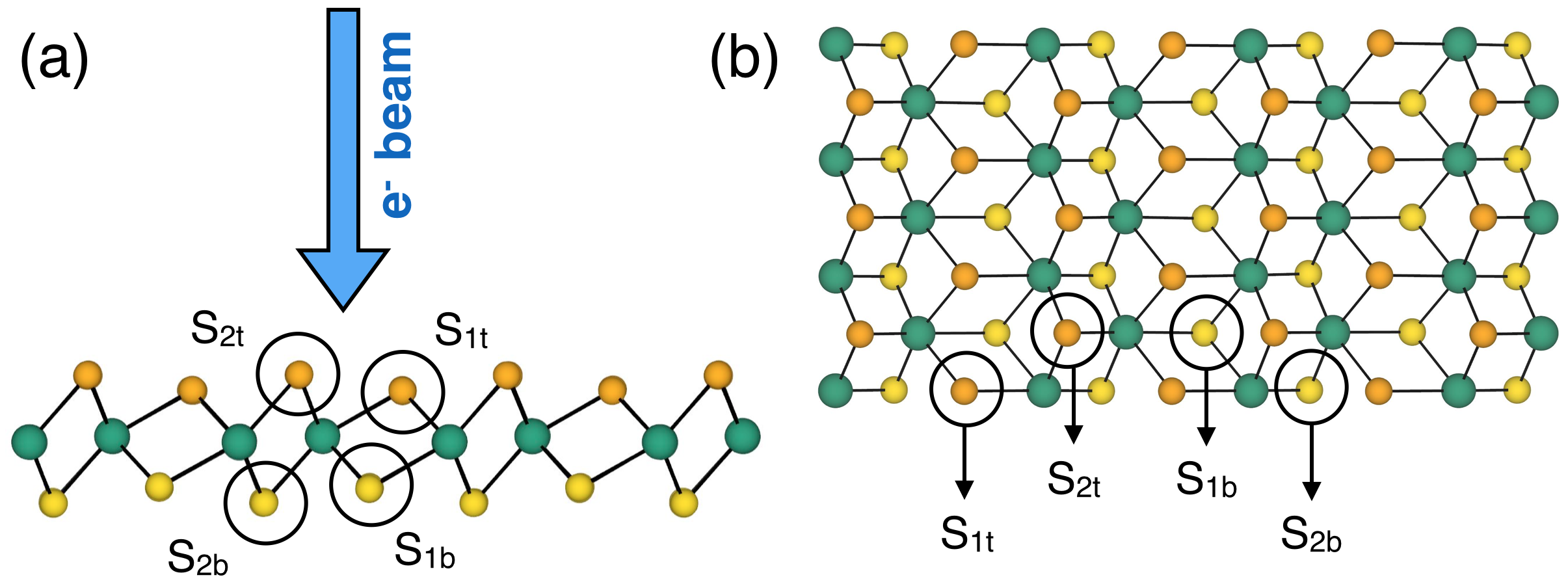}
  \caption{(a) Side-view and (b) top-view of the atomic structure of single-layer $1T'$-MoS$_2$. Molybdenum atoms are shown in green, while sulfur atoms belonging to the top and bottom chalcogen layers are orange and yellow, respectively. The direction of the electron beam and the labels of the nonequivalent sulfur lattice sites are  shown.  \label{Fig1}}
\end{figure}

\section{Methodology} 

The exposure of a solid to irradiation by high-energy electrons leads to beam-induced electron-electron and electron-nucleus scattering events. Primarily, the former result in electronic excitations and the latter in both phonon generation and possible ballistic displacement of atoms from their equilibrium positions. While in insulating media the effect of electronic excitations can be substantial, this is not the case of $1T'$-MoS$_2$ where they are expected to be quenched due to the metallic character of the system.  Furthermore, the energy scales of phonon modes (meV) are much smaller compared to those of atomic displacements (eV) and usually are not responsible for any damage in the irradiated samples. As a consequence, in this work we restrict our investigation to atoms displacements -- with a particular emphasis on knock-on damage -- that are the relevant events in the defect-engineering of materials\cite{bah99, krash10}.

Ballistic displacement of atoms takes place via elastic electron-nucleus scattering, a process that occurs  instantaneously ($\approx$10$^{-21}$ s) and results in an energy transfer from the incoming beam to the crystal, within a focus area that can be comparable to the interatomic distances\cite{krash10}. We model this process by assigning to one sulfur atom in  $1T'$-MoS$_2$ an initial momentum that corresponds to kinetic energy $T$ in the direction normal to the monolayer (see Fig.  \ref{Fig1}a). Such head-on collisions correspond to the largest transferred energy.\cite{bah99}  We then let the lattice evolve in time performing \emph{ab initio} molecular dynamics (AIMD) simulations in the microcanonical ensemble within the Born-Oppenheimer scheme. The displacement threshold energy $T_d$ corresponds to the minimum transferred energy necessary to permanently create a defect in the lattice upon irradiation. This methodology has been already successfully adopted to model the experimentally observed behavior of $2H$-MoS$_2$\cite{kom12} as well as carbon nanostructures\cite{yaz07} under irradiation.

Taking into account relativistic effects, the maximum energy $T_{max}$ that can be transferred from an incident beam of energy $E_{beam}$ to a nucleus of atomic mass $M$ as a result of the collision reads as\cite{garcia14}:
\begin{equation}
T_{max} = \frac{E_{beam}(E_{beam} + 2m_ec^2)}{E_{beam} + \frac{Mc^2}{2}(1 + \frac{m_e}{M})}
\end{equation}
with $m_e$ and $c$ being the mass of the electron at rest and the speed of light, respectively. When the maximum energy is equal to or larger than the displacement energy of a given defect (\emph{i.e.} $T_{max} \geq T_d$), then that defect is likely to form at the corresponding electron beam energy $E_{beam}$. This allows one to give an estimate of the electron energy at which certain structural defects form -- based on the determination of $T_d$ solely -- if a static lattice is assumed.

Furthermore, we investigate the relative stability of the sulfur vacancy defects through the determination of their formation energies $E_f$ defined as
\begin{equation}
E_f = E_{defect} - E_{pristine} + N \times \mu_{S} 
\end{equation} 
with $E_{defect}$ and $E_{pristine}$ being the total energies of our models with and without the defect, respectively, $N$ the number of removed sulfur atoms and $\mu_{S}$ their chemical potential, taken as the energy of the isolated sulfur atom. Notice that this is not the conventional thermodynamic formation energy (that should instead be defined with respect to elemental sulfur), but rather an operative definition that allows to quantify the energetics of sulfur atoms sputtering from the lattice to the gas phase, as it typically occurs during the electron beam irradiation.
 
Our periodic model system of single-layer $1T'$-MoS$_2$ consists of a rectangular 6 $\times$ 3  supercell containing 108 atoms, separated by its periodic replica by a vacuum region of 15~{\AA}. All our calculations are performed within the Density Functional Theory (DFT) framework\cite{jon15} relying on the generalized gradient approximation of Perdew, Burke and Ernzerhof\cite{per96}, as implemented in the Vienna \emph{ab initio} simulation package (VASP)\cite{kres93, kres96}. The electron-core interaction is described through the projector-augmented wave  (PAW) method\cite{kres97} while a plane-waves basis is used for valence electrons. Integration over the first Brillouin zone was carried out using a Monkhorst-Pack $k$-grid.   In molecular dynamics simulations, classical equations of motion were integrated with the Verlet algorithm and a timestep of 0.5 fs. Trajectories of 1--2 ps were found to be sufficiently long for our purposes. In order to accurately trace the $T_d$, for each lattice site we scan values of initial momenta ranging from 0 to 25~eV, with an energy resolution of 0.1~eV.  Kinetic energy cutoff was set to 300~eV and only the $\Gamma$ point was sampled during AIMD runs while geometry relaxations were performed with a larger cutoff of 500~eV, a denser mesh of 4 $\times$ 3 $\times$ 1 $k$-points and a tolerance on Hellmann-Feynman forces of 0.01~eV/{\AA}. Diffusion energy barriers were computed with the climbing-image nudged elastic band method\cite{henkel00} relaxing three intermediate images between the initial and final states until the residual force on each atoms drops down to 0.04~eV/{\AA}.

\begin{figure*}[]
  \begin{minipage}[]{0.75\textwidth}
    \includegraphics[width=\textwidth]{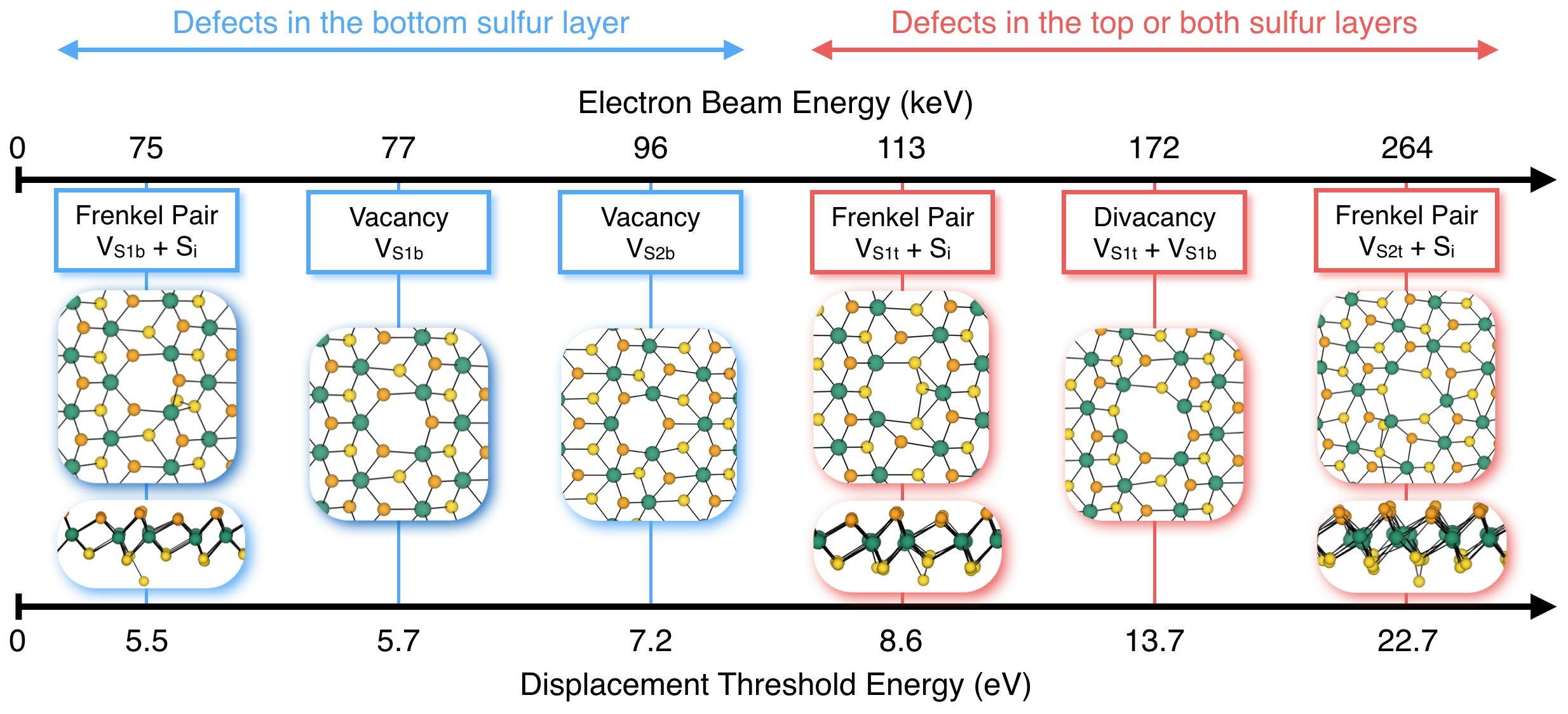}
  \end{minipage}\hfill
  \begin{minipage}[]{0.23\textwidth}
  \caption{Overview of the atomic defects observed in our molecular dynamics simulations that were performed on pristine single-layer $1T'$-MoS$_2$ together with their displacement threshold energies and corresponding electron beam energies. Lattice sites at which defects form are labelled according to Fig.  \ref{Fig1}. Blue (red) boxes correspond to defects forming at the bottom (top or both) chalcogen layer(s).  \label{Fig2}}
  \end{minipage}
\end{figure*}

\section{Results and Discussion}
\subsection{Electron irradiation of pristine $1T'$-MoS$_2$}

Two-dimensional single-layer $1T'$-MoS$_2$ consists of one layer of Mo atoms sandwiched between two layers of S atoms. As we pointed out in our previous work\cite{piz17}, in this lattice there exist two nonequivalent sulfur atoms per chalcogen layer. In order to distinguish these two positions, we will refer to the sulfur layer closer to (farther from) the molybdenum layer as S\textsubscript{1} (S\textsubscript{2}). In addition, we  label S\textsubscript{1t} or S\textsubscript{2t} and S\textsubscript{1b} or S\textsubscript{2b} the chalcogen atoms belonging to the top (t) or bottom (b) layer, respectively \footnote{Here and in the following we refer to top layer as the sulfur layer on the same side of  the electron beam source.}. This gives rise to four nonequivalent sulfur sites from the point of view of the head-on collision process, as we summarize for clarity in Fig. \ref{Fig1}.

As a first step, we consider  pristine $1T'$-MoS$_2$ and we determine the different events that take place when an increasing amount of energy is transferred from the electron beam to each of the four nonequivalent sulfur atoms. An overview of our results is presented in Fig. \ref{Fig2}. It is found that defects at the bottom chalcogen layer require lower threshold energies to form  (5.5 eV $< T_d < $  7.2 eV) compared to those at top or both chalcogen layers (8.6 eV $< T_d < $  22.7 eV). When a chalcogen atom in the top layer  (S\textsubscript{1t} and S\textsubscript{2t}) is displaced downwards, it is "stopped" and kicked back by the underlying Mo layer. This is not the case of S atom in the bottom layer (S\textsubscript{1b} and S\textsubscript{2b}), where atoms can therefore be sputtered more easily.

Specifically, our simulations suggest that a Frenkel pair forming at the S\textsubscript{1b} site -- in which one atom is removed from its equilibrium lattice site and is located in an interstitial position -- is characterized by a $T_d$ of 5.5 eV. However, when the energy transferred to the S\textsubscript{1b} atom is slightly increased by 0.2~eV, a V\textsubscript{S1b} defect forms. It is worth noting that the V\textsubscript{S1b} defect presents lower displacement energy than the V\textsubscript{S2b} by 1.5 eV. The reason for this can be traced back to the weaker Mo--S\textsubscript{1b} (or, equivalently, Mo--S\textsubscript{1t}) bond compared to the Mo--S\textsubscript{2b} (or Mo--S\textsubscript{2t}) bond, that can also be seen from the longer interatomic distance of the former (2.48 {\AA}) compared to the latter (2.39 {\AA})\cite{piz17}. For purpose of comparison, we also determine the displacement threshold energy of a single sulfur vacancy in the bottom layer of the semiconducting $2H$-MoS$_2$. In agreement with previous works\cite{kom12, garcia14}, we obtain 7.0~eV  suggesting that the $1T'$ phase is more susceptible to radiation-damage than the $2H$ phase, if the same electron energy is used.

\begin{figure*}[]
  \centering
  \includegraphics[width=1.9\columnwidth]{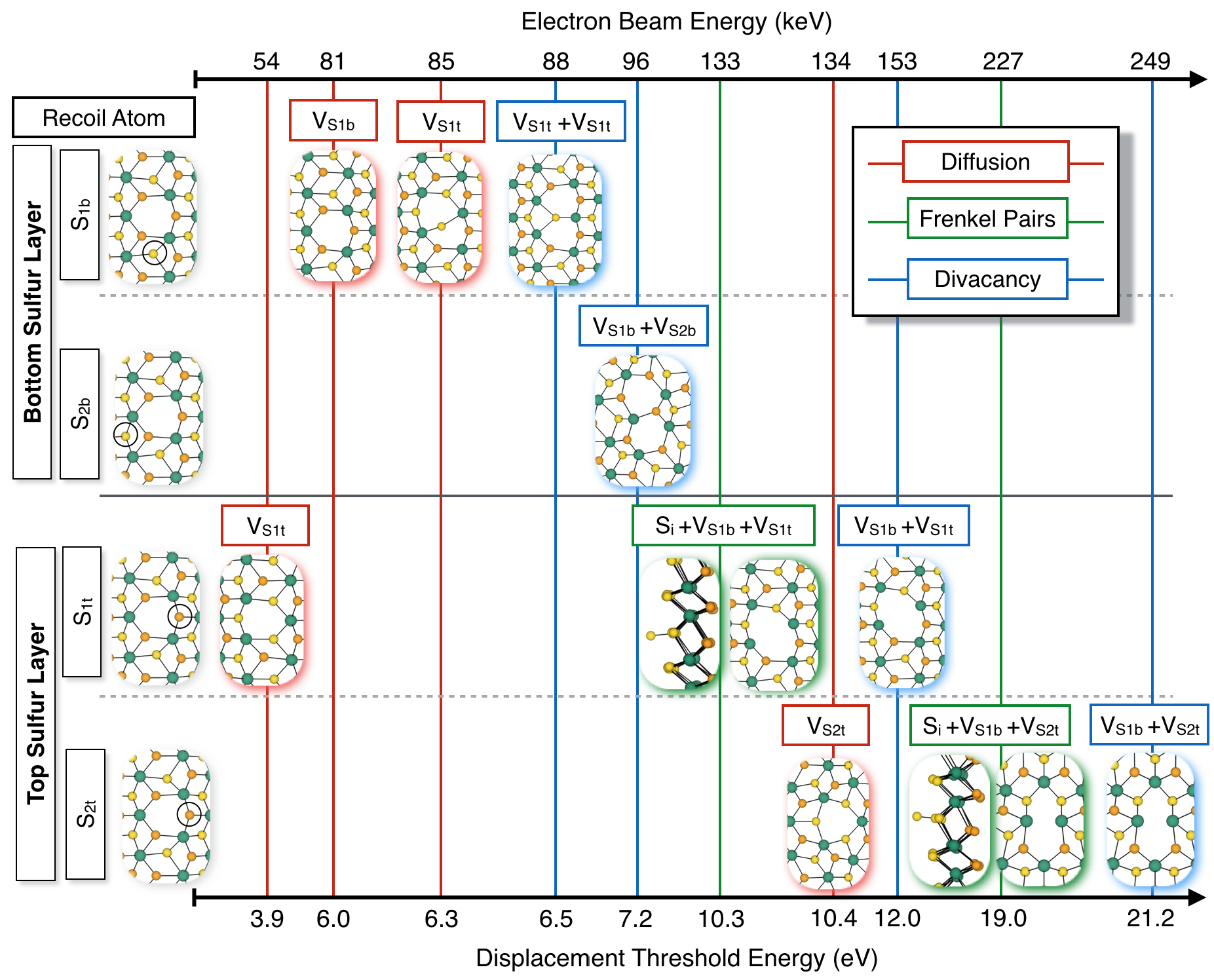}
  \caption{Overview of the atomic defects that emerge in the vicinity of a V\textsubscript{S1b} defect as a function of the displacement threshold energies and corresponding electron beam energies. For each row, the initial recoil atom  is shown in the black circle in the left column. According to the legend, red, green and blue boxes correspond to diffusion events, Frenkel pairs and divacancies formation, respectively. Lattice sites are labelled according to Fig. \ref{Fig1}.   \label{Fig3}}
\end{figure*}

It has been previously shown that in the $2H$ phase of single-layer transition metal dichalcogenides the displacement threshold energy to form a sulfur vacancy in the bottom layer is very similar to the formation energy calculated in the unrelaxed lattice\cite{kom12}.  This is indeed the case also for $1T'$-MoS$_2$, where we found unrelaxed formation energies of 5.97~eV and 6.98~eV for V\textsubscript{S1b} and V\textsubscript{S2b} defects, respectively, to be compared with their $T_d$ of 5.70~eV and 7.20~eV. This agreement deteriorates when the lattice is allowed to relax, lowering the formation energies to 5.35~eV and 6.42~eV.  We explain this in terms of time scales at which vacancy defects form in the irradiation process. Sulfur atom sputtering upon electron-nucleus scattering occurs within a few tens of femtoseconds, a timescale that allows only a little amount of energy to be dissipated at the neighboring lattice sites. Therefore, the energy to remove one sulfur atom from the frozen lattice can be considered a good approximation of the displacement threshold energy.  The difference between the formation energies assuming the frozen and relaxed lattices is 0.62~eV and~0.56 eV for V\textsubscript{S1b} and V\textsubscript{S2b} defects, respectively. This suggests that similar lattice relaxation takes place at the two different sulfur vacancies and that the larger formation energy of the V\textsubscript{S2b} defect compared to V\textsubscript{S1b} only stems from the stronger bond of S\textsubscript{2b}  compared to S\textsubscript{1b}, as we mentioned above.

Contrary to the bottom layer, we do not observe any single vacancy formation when sulfur atoms in the top layer undergo knock-on collision events. We found  that Frenkel pairs forming at S\textsubscript{1t} and S\textsubscript{2t} sites are characterized by threshold energies of 8.6 eV and 22.7 eV, again a remarkable difference between the two nonequivalent sulfur atoms.  A divacancy composed of V\textsubscript{S1t} and V\textsubscript{S2b} defects can form in the lattice when an energy of 13.7 eV is transferred to the  S\textsubscript{1t} lattice site. Our molecular dynamics simulations provide a detailed explanation of  the mechanism behind the origin of this divacancy, that turns out to involve the formation of a diatomic S$_2$ molecule. Specifically, the recoil atom S\textsubscript{1t} is knocked downards, being displaced by the electron beam, and further transfers momentum to the underlying S\textsubscript{1b} atom leading to the formation of the S$_2$ molecule that is eventually sputtered, leaving a divacancy in the lattice. Importantly, within the investigated range of $T_d$ we do not observe any vacancy defect involving the S\textsubscript{2t} sites, suggesting that this sulfur atom can be displaced only for $T_d >$ 25 eV.

In order to provide useful insights to experimentalists, we now turn our attention to the electron beam energies necessary to create the atomic defects described above, that we show in Fig. \ref{Fig2}. When the pristine lattice of $1T'$-MoS$_2$ is considered as a model of the crystal, it turns out that no defects should be produced below 75 keV: this is the larger beam energy at which imaging in the transmission electron microscope can be carried out  without leading to any substantial radiation-induced damage in the sample. Above this value, in fact, vacancies at the bottom layer should start forming. In the $2H$ crystalline phase, within the same static lattice approximation used here, vacancies are introduced when electron beam energy larger than 92 keV are employed. Such a  remarkable difference stems from the lower cohesive energy of $1T'$-MoS$_2$ compared to $2H$-MoS$_2$\cite{piz17}. Interestingly, the electron energy to form the V\textsubscript{S1b} defect (77 keV) is significantly lower compared to the one necessary to create the V\textsubscript{S2b} defect (96 keV). This suggests that an electron beam of energy equal to or larger than 77 keV but smaller than 96 keV can \emph{selectively} create vacancies at the S\textsubscript{1b} sites while keeping the S\textsubscript{2b} sublattice intact.  Furthermore, we note that only electron energies larger than 113 keV can lead to major damage in both sulfur layers. Importantly, we stress that the above mentioned values of electron beam energies do not account for lattice thermal motion and zero-point vibrations. As a consequence, the static beam energies mentioned above have to be viewed as an upper limit of the actual beam energies to induce knock-on damages in the sample, when a finite temperature is considered. In a recent work\cite{kom12}, it has been found that lattice vibrations at room temperature lowers the electron beam energy to remove a sulfur atom in 2$H$-MoS$_2$ by $\approx$10 keV, and we suggest that a similar estimate should hold also for irradiated 1$T'$-MoS$_2$.

\begin{figure}[]
  \centering
  \includegraphics[width=0.9\columnwidth]{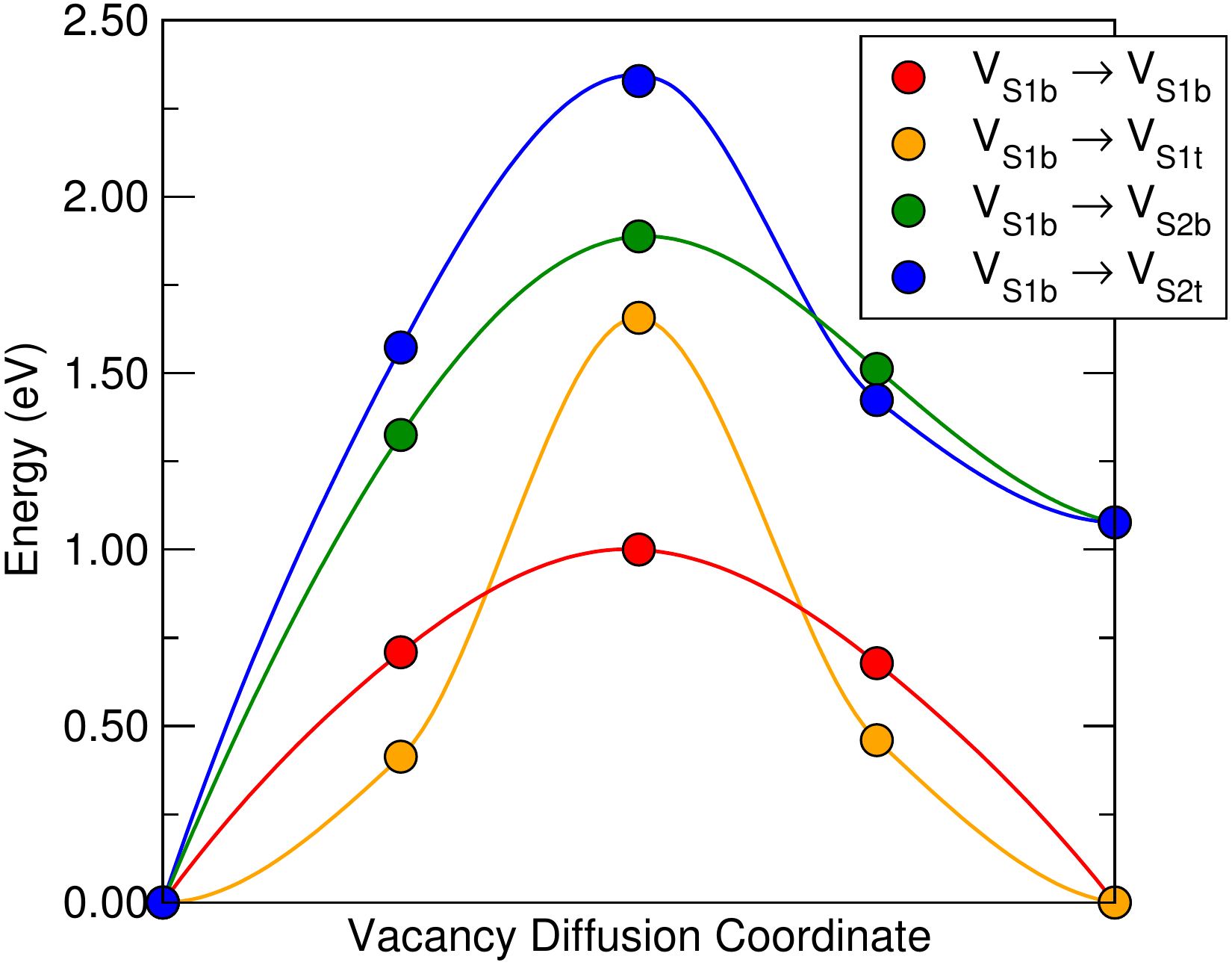}
  \caption{Energy profiles for the diffusion of the stable V\textsubscript{S1b} defect into different positions in single-layer $1T'$-MoS$_2$ calculated using the climbing-image nudged elastic band method. Lattice sites at which the vacancy defects form are labelled according to Fig. \ref{Fig1}. Calculated points are interpolated with splines to guide the eye.  \label{Fig4}}
\end{figure}

\subsection{Electron irradiation of defective $1T'$-MoS$_2$}

Next, we investigate  $1T'$-MoS$_2$ under the electron beam in the vicinity of a sulfur vacancy. Specifically, we introduce the V\textsubscript{S1b} defect in the otherwise pristine monolayer. The reason for this choice is twofold. On one hand, the V\textsubscript{S1b} impurity presents the lowest displacement threshold energy and therefore it can readily form during irradiation (Fig. \ref{Fig2}). On the other hand, in our previous work\cite{piz17} we have found that this defect is characterized by a very low formation energies -- $\approx$1 eV lower than the frequently observed sulfur vacancy in the stable $2H$ phase of MoS$_2$ -- thereby suggesting that it is likely to be the most common native impurity when thermodynamic equilibrium prevails\cite{piz17}.  In the following,  we focus on the response of each of the four nonequivalent sulfur atoms closest to the V\textsubscript{S1b} defect to the electron beam. Our results are presented in Fig. \ref{Fig3}.  Depending on the magnitude of the transferred momentum during the irradiation, two different types of events take place around the impurity, namely (i) the sulfur vacancy diffusion and (ii) the Frenkel pair formation followed by the formation of divacancy defects.

Diffusion events of atoms in the bottom chalcogen layer are very likely to occur around the vacancy defect under irradiation. In particular, we found that a transferred momentum as low as 3.9 eV can result in diffusion of V\textsubscript{S1b} defect in the out-of-plane direction, leading to the V\textsubscript{S1t} defect. This suggests  that an electron beam energy of 54 keV, though not sufficient for inducing any atom sputtering, is nevertheless large enough to promote the diffusion of vacancy defects, if present in the monolayer. Furthermore, when the S\textsubscript{1b} lattice site acts as a recoil atom, both in-plane and out-of-plane diffusion paths are observed with corresponding $T_d$ of 6.0 and 6.3 eV (or, equivalently, electron beam energies of 81 keV and 85 keV), respectively.  The mobility of the sulfur vacancy V\textsubscript{S2t} belonging to the top sulfur layer is observed only at larger transferred momentum of 10.4 eV (or beam energy of 134 keV) to in-plane diffusion to a neighboring equivalent site. Overall, diffusion processes involving S\textsubscript{1b} and S\textsubscript{1t} sites are achieved at lower beam energies than those involving the S\textsubscript{2t}, further  highlighting the different bond strength between the nonequivalent sulfur atoms in the $1T'$-MoS$_2$ lattice.

We extend our investigation of the mobility of the stable V\textsubscript{S1b} defect by computing its energy barriers to diffuse to all possible neighboring sites, \emph{i.e.} the S\textsubscript{1b}, S\textsubscript{1t}, S\textsubscript{2b} and S\textsubscript{2t} sites,  as shown in Fig. \ref{Fig4}. It is found that the lowest barrier path involves the in-plane diffusion of the V\textsubscript{S1b} defect to an equivalent S\textsubscript{1b} site with a barrier equals to 1.00 eV, while the out-of-plane diffusion to the S\textsubscript{1t} site requires  a larger barrier of 1.66 eV to occur. These latter paths are likely to take place at room temperature and are overall the most energetically favorable ones. This is consistent with our molecular dynamics simulations reported in Fig. \ref{Fig3}, where we found that the diffusion along the above mentioned paths occurs at the lowest electron beam energies.  
On the other hand, diffusion processes of the  V\textsubscript{S1b} defect to the S\textsubscript{2b} and S\textsubscript{2t} sites are characterized by larger energy barriers of 1.89 eV and 2.33 eV, respectively. While we observe the latter diffusive event in AIMD simulations at a beam energy of 134 keV, we do not observe the former, even though the barrier is lower by 0.44 eV. We suggest that the reason for this is possibly due to the direction of the incoming beam: because the electron beam is incident along the direction normal to the monolayer (head-on collision), it is unlikely to observe in-plane diffusions requiring larger activation barriers to take place.

\begin{figure}
  \centering
  \includegraphics[width=1\columnwidth]{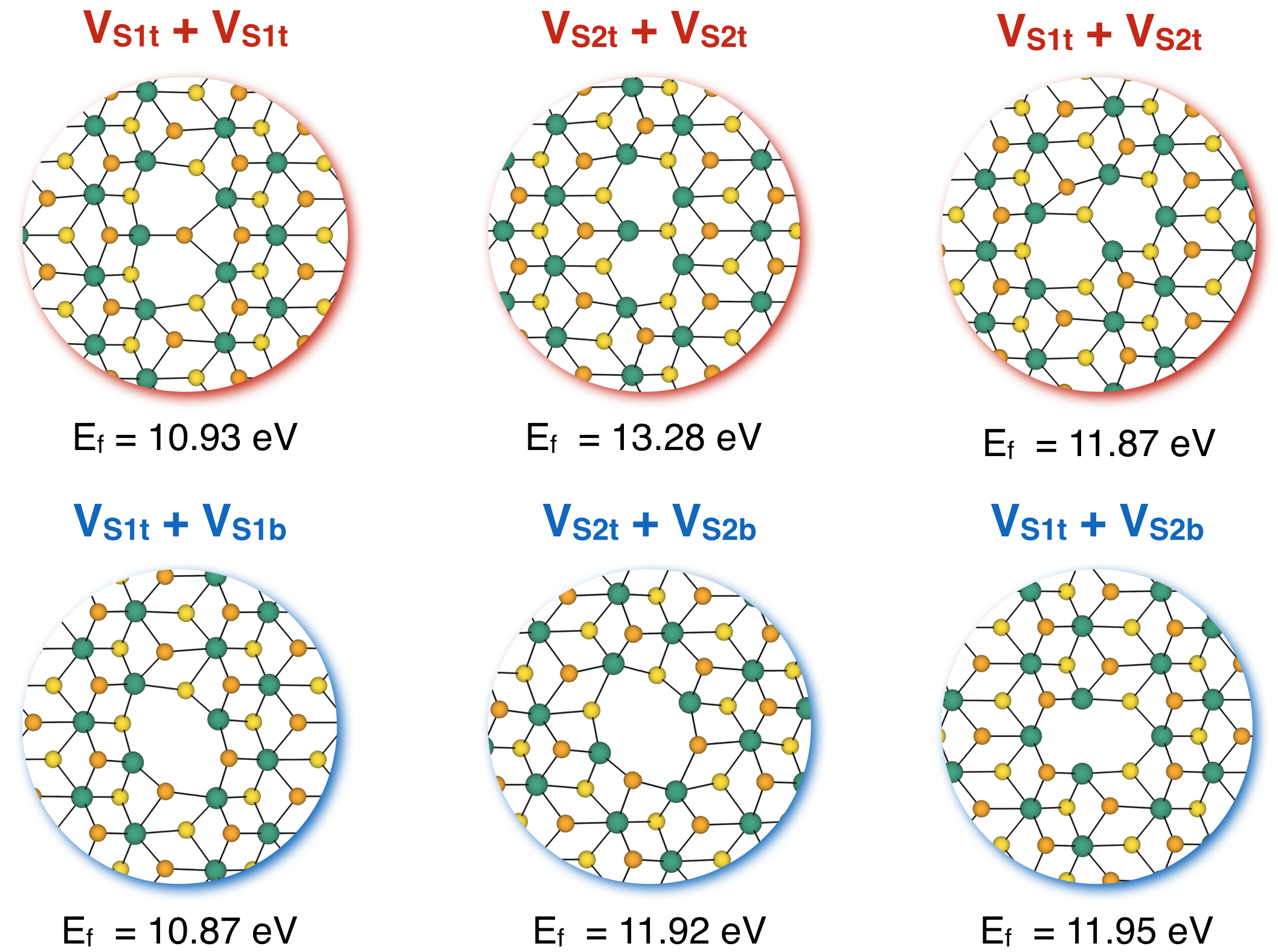}
  \caption{Atomic structures and formation energies ($E_f$) of all possible sulfur divacancy defect configurations in $1T'$-MoS$_2. $ Blue (red) color indicates divacancies forming at the same (opposite) chalcogen layers. Lattice sites are labeled according to Fig. \ref{Fig1}  \label{Fig5}}
\end{figure}

It is an interesting question whether a single sulfur vacancy favors or not the formation of a second vacancy at its neighboring sites under electron irradiation. In order to address this issue, we identify the displacement threshold energies of the S\textsubscript{1b}, S\textsubscript{2b}, S\textsubscript{1t} and S\textsubscript{2t} sites closest to the vacant site (Fig. \ref{Fig3}). 

Concerning the bottom chalcogen layer, we found that divacancy defects are characterized by $T_d$ of 6.5 and 7.2 eV, depending on whether the S\textsubscript{1b} or S\textsubscript{2b} site is considered as a recoil atom, respectively. When comparing those threshold energies with the ones of the single vacancy reported in Fig. \ref{Fig2}, it turns out that the sulfur vacancy does not play any significant role in affecting the threshold energies (or electron beam energy) of the nearest S\textsubscript{2b} lattice site -- that thus remain unchanged --  while it slightly increases the $T_d$ of S\textsubscript{1b} sites in the bottom layer by 0.7 eV.  Hence, the presence of V\textsubscript{S1b} does not favor the formation of a second vacancy in its neighborhood in the bottom layer. 

Then, we inspect the response of the  S\textsubscript{1t} and S\textsubscript{2t} sites closest to the  V\textsubscript{S1b} defect. First, we observe that similarly to the pristine lattice, the electron beam energy necessary to create defects in the top layer of defective lattice is larger than in the bottom layer. Second, Frenkel pairs form at threshold energies of 10.3 eV and 19.0 eV  at S\textsubscript{1t} and S\textsubscript{2t} sites, respectively, and precede in energy the formation of divacancies by  $\approx$2 eV.  Furthermore, we note that such divacancies have lower threshold energies (12.0 eV and 21.2 eV) than the corresponding ones forming in the same layer of pristine lattice (13.7 eV and no defect formation observed up to 25~eV). We also observe that V\textsubscript{S1b}+V\textsubscript{S1t} and V\textsubscript{S1b}+V\textsubscript{S2t} divacancies should form under beams of energies at least equal to 134 keV and 249 keV, respectively. 

Summarizing, our simulations suggest that the presence of the V\textsubscript{S1b} defect lowers the displacement threshold energies for the formation of a second vacancy only in the top layer but not in the bottom one.  In the case of pristine lattice, the bottom layer is intact and therefore it stops the recoil atom when displaced downwards upon the irradiation. However, when V\textsubscript{S1b} is introduced, this stopping effect is of minor extent because of the missing atom, thereby lowering the displacement threshold energies of the sulfur atoms in the top layer. 

As a final subject of investigation, we focus on the stability of divacancy defects in $1T^\prime$-MoS$_2$ through the determination of their formation energies (see Fig. \ref{Fig5}). We stress that formation energies account for the stability of defect under thermodynamic equilibrium while electron beam irradiation is a non-equilibrium process. Nevertheless, it is possible to draw some qualitative connection between $E_f$ and $T_d$.
Specifically, it is found that the most stable divacancies configurations are those involving the S\textsubscript{1t} and/or S\textsubscript{1b}  lattice sites, namely V\textsubscript{S1t}+V\textsubscript{S1t} and V\textsubscript{S1t}+V\textsubscript{S1b} \footnote{Note that in the absence of the electron beam -- as in the case of the calculation of formation energies -- due to symmetry it turns out that: V\textsubscript{S1t}+V\textsubscript{S1t} = V\textsubscript{S1b}+V\textsubscript{S1b}, V\textsubscript{S2t}+V\textsubscript{S2t} = V\textsubscript{S2b}+V\textsubscript{S2b}, V\textsubscript{S1t}+V\textsubscript{S2t} = V\textsubscript{S1b}+V\textsubscript{S2b}, V\textsubscript{S1t}+V\textsubscript{S1b} = V\textsubscript{S1b}+V\textsubscript{S1t}, V\textsubscript{S2t}+V\textsubscript{S2b} = V\textsubscript{S2b}+V\textsubscript{S2t} and V\textsubscript{S1t}+V\textsubscript{S2b} = V\textsubscript{S1b}+V\textsubscript{S2t}. }. These configurations also emerge from our molecular dynamics simulations at electron beam energies of 88 keV and 134 keV, respectively.  On the other hand, the V\textsubscript{S2t}+V\textsubscript{S2t} divacancy presents the largest formation energies, and consistently it does not appear as a result of the AIMD calculations. Finally, we note that V\textsubscript{S1t}+V\textsubscript{S2t}, V\textsubscript{S1t}+V\textsubscript{S2b} and V\textsubscript{S2t}+V\textsubscript{S2b} divacancies show very similar formation energies that only differ by few tens meV. Despite this, only the former configurations are observed as a result of the dynamics while the latter is not, thereby indicating a competition between equilibrium and non-equilibrium processes during the electron beam irradiation. 

\section{Summary and Conclusions} 

In summary, motivated by the recent progress in defect-engineering of transition metal dichalcogenides in the electron microscope, we have investigated for the first time the response of $1T'$-MoS$_2$ to the electron irradiation using \emph{ab initio} molecular dynamics simulations.  

We have found that an electron beam energy below 75 keV has to be used in order to perform TEM imaging on the $1T'$-MoS$_2$ samples without leading to any substantial knock-on damage. Sulfur atoms belonging to the bottom layer present lower displacement threshold energies compared to those in the top layer. As a consequence, an electron beam of energy up to 113 keV can selectively create vacancy defects in the bottom layer while preserving  the top one intact. Furthermore, in the bottom layer the sulfur atoms closer to the Mo plane have the lowest displacement threshold energies among all lattice sites. This implies that, when a proper tuning of the beam energy (77 keV $ < E_{beam} < $ 96 keV) is adopted, they can be selectively sputtered from the lattice,  leading to the formation of ordered  defects. 

In addition, sulfur vacancies are mobile under the electron beam. Depending on the specific path considered, calculated energy barriers to diffusion range from 1.00 to 2.33 eV, suggesting that vacancy defects can be mobile also at room temperature.  We also found that the presence of a single vacancy lowers the displacement threshold energies of  sulfur atoms at certain neighboring sites. Finally, we have discussed the most stable configurations of double vacancies when thermodynamic equilibrium prevails and provided a comparison with those that are expected to be observed upon electron irradiation. 

Overall, our simulations suggest that the metastable $1T'$ phase of single-layer MoS$_2$ is more susceptible to knock-on damage compared to the thermodynamically stable $2H$ when the same electron beam energy is adopted.  Recently, it has been theoretically shown that the introduction of sulfur vacancy defects and  the accompanying lattice strain are able to reduce the difference in energy between the 2$H$ phase and the 1$T'$ polymorphs of MoS$_2$\cite{kretsch17}. As vacancies can be created easier in 1$T'$-MoS$_2$ compared to 2$H$-MoS$_2$, our results suggest that, once the 1$T'$ phase is formed, the subsequent introduction of defects through electron irradiation should stabilize it over the otherwise thermodynamically stable 2$H$ phase.

In conclusion, our findings provide useful insights to the imaging of $1T'$-MoS$_2$ in the electron microscope together with important guidelines for the defect engineering. 

\begin{acknowledgments} 
M. P. gratefully acknowledges Hannu-Pekka Komsa  at Aalto University for fruitful discussions and Vamshi M. Katukuri at EPFL for technical assistance. This work was financially supported by the Swiss National Science Foundation (Grant No. 20021\_162612).  All calculations were performed at the Swiss National Supercomputing Centre (CSCS) under the project s675.
\end{acknowledgments} 

\bibliography{References}

\end{document}